\documentclass[journal]{IEEEtran}
\ifCLASSINFOpdf
\else
\fi

\usepackage[english]{babel}
\usepackage[autostyle, english = british]{csquotes} %british single quote, American double quotes
\MakeOuterQuote{"}

\usepackage{epigraph}
\newcommand{\footurl}[1]{\footnote{\url{#1}}}

\usepackage{graphicx} % Required for inserting images
\usepackage{setspace}

\usepackage{physics}

\usepackage{todonotes}
\usepackage{url}

\newcommand{\tild}[0]{~}

\newcommand{\new}[1]{\textcolor{black}{#1}}

\begin{document}
\title{\new{From Quantum Mechanics to\\Quantum Software Engineering:\\A Historical Review}}

\author{Giuseppe Bisicchia, Jose Garcia-Alonso, Juan M. Murillo, and Antonio Brogi% <-this % stops a space
\thanks{G. Bisicchia and A. Brogi are with the Department of Computer Science, University of Pisa (Pisa, Italy).}% <-this % stops a space
\thanks{J. Garcia-Alonso and J. M. Murillo are with Quercus Software Engineering Group, University of Extremadura (Cáceres, Spain).}% <-this % stops a space
\thanks{Corresponding author: giuseppe.bisicchia@phd.unipi.it}
\thanks{Work partly supported by projects: \textit{hOlistic Sustainable Management of distributed softWARE systems (OSMWARE)}, PRA\_2022\_64, funded by the University of Pisa; \textit{PID2021-1240454OB-C31}, funded by the Spanish Ministry of Science, Innovation and Universities; \textit{GR21133}, funded by the Regional Ministry of Economy, Science and Digital Agenda of the Regional Government of Extremadura.}
}


\markboth{G. Bisicchia \textit{et al.}}%
{Bisicchia \MakeLowercase{\textit{et al.}}: From Quantum Mechanics to Quantum Software Engineering}



\maketitle

\begin{abstract}
Victor Hugo's timeless observation, "Nothing is more powerful than an idea whose time has come", resonates today as Quantum Computing, once only a dream of a physicist, stands at the threshold of reality with the potential to revolutionise the world. To comprehend the surge of attention it commands today, one must delve into the motivations that birthed and nurtured Quantum Computing. While the past of Quantum Computing provides insights into the present, the future could unfold through the lens of Quantum Software Engineering. Quantum Software Engineering, guided by its principles and methodologies investigates the most effective ways to interact with Quantum Computers to unlock their true potential and usher in a new era of possibilities. To gain insight into the present landscape and anticipate the trajectory of Quantum Computing and Quantum Software Engineering, this paper embarks on a journey through their evolution and outlines potential directions for future research. By doing so, we aim to equip readers (ideally software engineers and computer scientists not necessarily with quantum expertise) with the insights necessary to navigate the ever-evolving landscape of Quantum Computing and anticipate the trajectories that lie ahead.
\end{abstract}

%Victor Hugo's timeless observation, "Nothing is more powerful than an idea whose time has come", resonates today as Quantum Computing, once only a dream of a physicist, stands at the threshold of reality with the potential to revolutionise the world. To truly comprehend the surge of attention it commands today, one must delve into the motivations that birthed and nurtured Quantum Computing. While the past of Quantum Computing provides insights into the present, the future could unfold through the lens of Quantum Software Engineering.
%
%Drawing parallels with Classical Computing, Quantum Software Engineering, guided by its principles, methodologies, and tools, investigates the most effective ways to interact with Quantum Computers. This endeavour aims to unlock the true potential of Quantum Computing and usher in a new era of possibilities.
%
%\new{To gain insight into the present landscape and anticipate the trajectory of Quantum Computing and Quantum Software Engineering, this paper embarks on a journey through their evolution. Beginning with their foundations in Quantum Mechanics, we trace the historical development of Quantum Computing and delve into the emergence of Quantum Software Engineering. By examining and discussing the underlying motivations that shaped this burgeoning field, we shed light on its development. Finally, we outline potential directions for future research in Quantum Software Engineering.}


% \fi
\IEEEpeerreviewmaketitle

\section{A Brief History of Quantum Computing}
\label{sec:intro}

% \epi{The most incomprehensible thing about the universe is that it is comprehensible.}{Albert Einstein}
% \epi{There is no a priori reason why physical laws should respect the limitations of the mathematical processes we call ‘algorithms’.}{David Deutsch}
% %\vspace{-10mm}\epi{It is often claimed that every ‘reasonable’ physical model for computation, is equivalent to Turing’s. But this is not so; there is no a priori reason why physical laws should respect the limitations of the mathematical processes we call ‘algorithms’.}{David Deutsch}
% \epi{Computers aren’t just human inventions. They are a fundamental feature of the universe, the answer to a simple and profound question about how the universe works.}{Michael Nielsen
% }
% %\epi{The reason why we find it possible to construct, say, electronic calculators, and indeed why we can perform mental arithmetic, cannot be found in mathematics or logic. The reason is that the laws of physics ‘happen to’ permit the existence of physical models for the operations of arithmetic such as addition, subtraction and multiplication.}{David Deutsch}

% \vspace{5mm}

\textit{"What kind of computer are we going to use to simulate physics?"} It was the Nobel laureate Richard Feynman who raised this question in his visionary talk\tild\cite{Feynman1982}, beginning the history of Quantum Computing\footnote{To be fair, already a few years earlier, in 1980, mathematician Yuri Manin pointed out the problems in simulating quantum systems with classical computers\tild\cite{manin1980computable}. Independently, between 1980 and 1982, the physicist Paul Benioff proposed a quantum mechanical model of Turing machines\tild\cite{benioff1980computer,benioff1982quantum}, a topic also discussed by Feynman in 1985\tild\cite{feynman1985quantum}.}.

This question is rooted in a series of crises and revolutions\tild\cite{planck1901ueber,einstein1905erzeugung,bohr1913constitution,heisenberg1925quantentheoretis,schrodinger1926quantisierung,born1926quantenmechanik} that shook the world of physics to its foundations between 1900 and 1925. The result of that period of turmoil was a new fundamental theory of physics which describes the behaviour of Nature at subatomic levels: Quantum Mechanics\tild\cite{dirac1981principles,von2018mathematical,pauli2012general}. %Quantum Mechanics is a theoretical framework for the construction of physical theories in which physical systems are described by complex-valued \textit{state vectors}.

Manin and Feynman's concerns\tild\cite{Feynman1982,manin1980computable,feynman1985quantum} on the simulation of physical systems were primarily about the difficulties of modelling \textit{quantum} systems. In such systems, the number of variables required to represent them increases exponentially with their complexity and with the number of particles involved\footnote{On the other hand, the number of variables to describe classical physical systems grows only linearly with their complexity.}.

Following these considerations, in 1985, physicist David Deutsch suggested, in his seminal work\tild\cite{deutsch1985quantum}, a deeper connection between computing and physics, stating a stronger "physical version" of the \textit{Church-Turing thesis}\footnote{Here is reported a version, from\tild\cite{deutsch1985quantum}, for completeness: \textit{"Every function which would naturally be regarded as computable can be computed by the universal Turing machine"}.}. Such thesis, also called the \textit{Church–Turing–Deutsch principle} states that:

\begin{quotation}
\noindent
\textit{Every finitely realizable physical system can be perfectly simulated by a universal model computing machine operating by \mbox{finite means.}}
\end{quotation}
\smallskip

With his physical interpretation of the \textit{Church-Turing thesis}, Deutsch brought attention to an often neglected fact about computation. Every algorithm is actually performed by a physical system, whether it be an electronic calculator, a mechanical apparatus or a human being. Thus, computation is ultimately a physical process and, hence, a \textit{universal} computer (that is a physical system too) must be able to simulate the dynamics of every possible physical system.

%think physically about computation

% Any physical theory, not just quantum mechanics, may be used as the basis for a theory of information processing and communication. 

The consequences of the physics revolution in the early 20\textsuperscript{th} century, however, led scientists to postulate that the fundamental nature of physics is ultimately quantum mechanical. Unfortunately, classical systems seem to be ineffective in efficiently simulating quantum mechanical systems\tild\cite{Feynman1982,manin1980computable}.
Deutsch was then naturally led to propose a universal computing device based on the principles of quantum mechanics\tild\cite{deutsch1985quantum}, so as to overcome the limitations of classical computers: the quantum computer was born.

Pretty soon, the potential of quantum computers began to be, as Deutsch surmised, far more impactful than just simulating physical systems. In 1992, David Deutsch, in collaboration with Richard Jozsa, formulated a problem that, even if of little practical interest, can be solved more efficiently by quantum devices than by any classical or stochastic algorithm\tild\cite{deutsch1992rapid}. Shortly afterwards, in 1993, Ethan Bernstein and Umesh Vazirani proposed another problem, showing the advantage of quantum devices over classical ones even when small errors are allowed\tild\cite{bernstein1993quantum}. In the same work, Bernstein and Vazirani designed a quantum version of the Fourier transform\footnote{\new{The Quantum Fourier Transform serves as the quantum counterpart to the Discrete Fourier Transform, offering an efficient quantum algorithm for implementing Fourier Transform operations. Crucial in various quantum algorithms, it plays a fundamental role in extracting and translating purely quantum information stored within qubits into classically measurable outcomes.}}\tild\cite{nussbaumer1982fast}. In 1994, leveraging the quantum Fourier transform and the work of Daniel Simon\tild\cite{simon1994power}, who showed that a quantum computer could find the period of a function with an exponential speedup, Peter Shor presented an efficient quantum algorithm for computing discrete logarithms. Only a few days later, Shor formulated an efficient quantum algorithm also for factoring large numbers\tild\cite{shor1994algorithms}. Both problems are believed to be intractable on classical computers, and thereby commonly used in cryptographic protocols\tild\cite{mccurley1990discrete,buchmann2004introduction}. Just two years later, Seth Lloyd proved that quantum computers could simulate quantum systems without the exponential overhead present in classical simulations, confirming Feynman's 1982 conjecture\tild\cite{lloyd1996universal}. In the same year, Lov Grover presented a quantum algorithm achieving an optimal quadratic speedup for unstructured search\tild\cite{grover1996fast}. Shor and Grover's breakthroughs gave a strong impetus to the research on quantum algorithms, demonstrating the existence of \textit{useful} problems that benefit from a quantum speedup. 

Meanwhile, research into working quantum computers also began to take its first steps. In 1993, Seth Lloyd proposed a method for building a potentially realisable quantum computer through pulsed quantum arrays, i.e., arrays of weakly coupled quantum systems subjected to a sequence of electromagnetic pulses of specific lengths and frequencies\tild\cite{lloyd1993potentially}. Not long afterwards, in 1995, Juan Cirac and Peter Zoller suggested an implementation of a quantum computer employing cold ionised atoms confined in electric potential traps and interacting with laser beams\tild\cite{cirac1995quantum}. Following the developments in the field, just one year later, David DiVincenzo formalised five minimal requirements for creating a working quantum computer. Such criteria include the availability of scalable qubits\footnote{\new{A \textit{"qubit"} is the computational unit of a Quantum Computer (as opposed to the classical \textit{bit}). A qubit state can be 0, 1 or in a \textit{superposition} (i.e., linear combination) of both. In the latter, when measured it will be only 0 or 1, with different probabilities according to its superposition.}} highly isolated from the external environment, the ability to initialise, manipulate and entangle their state and to "strongly"\footnote{With "strong" measurement it is meant a measure capable of collapsing the state of a qubit.} measure the state of each qubit\tild\cite{divincenzo1997topics}. A further milestone was set by Yasunobu Nakamura \textit{et al.}, between 1991 and 2001, who built a working, controllable superconducting qubit through a Josephson junction\tild\cite{nakamura1999coherent,nakamura2001rabi}.

In those years, however, a shadow threatened and questioned the very foundations of Quantum Computing\tild\cite{landauer1995quantum}. The \textit{decoherence} menaced to dash any hopes of having actually usable quantum computers\tild\cite{unruh1995maintaining}. Decoherence is the phenomenon that, under typical conditions, prevents complex many-particle quantum systems from exhibiting quantum behaviour for a long time, stranding the dream of a quantum computer with no way out\tild\cite{haroche1996quantum}. It was Shor again who gave hope and new life to the field. Shor demonstrated in 1995 how it was possible to reduce the destructive effects of decoherence through the quantum analogue of error-correcting codes\tild\cite{shor1995scheme} and fault-tolerant methods for executing reliable quantum computations on noisy quantum computers\tild\cite{shor1996fault}. The work of Shor and others\tild\cite{aharonov1997fault,knill1998resilient,kitaev1997quantum} thus confirmed that it is possible, at least in principle, to suppress the error rate of a quantum computer to arbitrarily low levels, thanks to error correction schemes and as long as the error rate is below a certain threshold\footnote{The assumptions made regarding the computational capability have a significant impact on the precise value of the threshold, but it is considered in the range $10^{-4} - 10^{-6}$\tild\cite{Nielsen2012}.}, this is the so-called \textit{threshold theorem}\tild\cite{preskill1998fault}.

%\section{Today}

Significant developments have been made since those first steps in both quantum software and hardware\tild\cite{Nielsen2012,preskill2023quantum}. In 2011, the first ever commercially available quantum computer was presented, and sold, by D-Wave\tild\cite{merali2011first}. It was D-Wave One, a 128-qubit quantum annealer\footnote{\new{A \textit{quantum annealer} is a specialised form of quantum computer. Unlike universal quantum computers, quantum annealers are non-Turing complete devices tailored specifically for solving optimisation problems as energy minimisation problems. Roughly speaking, quantum annealers ensure that each qubit eventually settles into a classical state that reflects the minimum energy configuration of the problem\tild\cite{annealing}.}}\tild\cite{Johnson2011}. In 2016, IBM put online their 5-qubit, gate-based\footnote{\new{Unlike quantum annealers, \textit{gate-based} quantum computers are universal computing machines. A gate-based quantum computer operates by manipulating qubits through the quantum analogue of classical logical gates\tild\cite{gatebased}.}}, superconducting quantum computer, making quantum computing publicly available for the first time, through the cloud\tild\cite{https://doi.org/10.48550/arxiv.1610.06980,Alsina2016}. In 2018, the first commercial quantum computer employing trapped ions was launched by IonQ\tild\cite{Ball2019}. Just one year later, Google claimed the achievement of quantum supremacy\footnote{Quantum Supremacy is the goal to \textit{"perform tasks with controlled quantum systems going beyond what can be achieved with ordinary digital computers"}\tild\cite{https://doi.org/10.48550/arxiv.1203.5813}.} with their 54-qubit, superconducting processor \textit{"Sycamore"}\tild\cite{Arute2019}. However, some doubts arose shortly afterwards\tild\cite{Liu2021,Bulmer2022,McCormick2022} and eventually classical devices beat Google's result\tild\cite{Pan2022}. The last current milestone in the quantum race was set in 2023 by IBM, which announced evidence for the utility of quantum computing even with noisy hardware, showing how it is possible to produce reliable results even without fault-tolerant quantum computers and at a scale beyond brute-force classical computation\tild\cite{Kim2023}. Also in this case, though, the scientific community does not entirely agree\tild\cite{https://doi.org/10.48550/arxiv.2306.14887,https://doi.org/10.48550/arxiv.2306.16372,https://doi.org/10.48550/arxiv.2306.15970,https://doi.org/10.48550/arxiv.2306.17839}.

%IBM q system one, first IBM commercial device 2019

Nevertheless, even though the supremacy and utility of quantum computers have not yet been established beyond a shadow of a doubt, there is no denying that we are now at the gates of a new era\tild\cite{knight2018serious,ibm2021quantum,carleton2021architecting}. Indeed, even if quantum and classical computers feature the same computational power\tild\cite{deutsch1985quantum}, i.e., they can solve the same class of problems, it is believed (and some evidence began to \new{arise)\tild\cite{daley2022practical,madsen2022quantum,zhong2020quantum,wu2021strong}}, that quantum computers can solve some problems asymptotically faster than what it is possible just with classical resources\tild\cite{Simon1997}. In fact, increasingly cutting-edge
applications are emerging, promising to revolutionise numerous industries and sectors and with a potentially immeasurable impact on society\tild\cite{lopez2019quantum}. Among the most researched areas are medicine, chemistry and pharmacy, biology and agriculture, engineering, energy and logistics, economy and finance, meteorology, manufacturing and cybersecurity\tild\cite{bayerstadler2021industry,bova2021commercial}.

\section{The Dawn of Quantum Software Engineering}

Despite the great and fast progress being made in Quantum Computing, current quantum computers cannot scale beyond dimensions of a few tens (or in the best cases hundreds) of qubits. At the same time, quantum devices are still very sensitive to external interference (noise), which can easily disrupt an ongoing computation. Due to such limitation, current quantum computers are usually referred to as \textit{Noisy Intermediate-Scale Quantum} (NISQ) devices\tild\cite{preskill2018quantum}, highlighting their capacity to execute only Quantum programs featuring a small number of qubits and consecutive steps\tild\cite{nisqalgo,leymann2020bitter,gemeinhardt2021quantum}.

However, this is not the first time in history that computer scientists have had to face such limitations on computing devices. Several authors, indeed, compare the current quantum computing landscape to that of classical computing during the 60s and that a similar development should be followed\tild\cite{moguel2020roadmap,serrano2022quantum,talaveramanifesto}.

In such a roadmap, Quantum Software Engineering has a primary role \textit{"to exploit the full potential of commercial quantum computer hardware, once it arrives"}\tild\cite{grandchallenge}. Quantum Software Engineering will be, indeed, also necessary to define the best quantum software development and application management lifecycles. They will enable to coherently employ and operate the increasing amount of quantum methodologies and tools, proposed to solve problems nowadays present in all the development and management phases\tild\cite{zhao2020quantum,weder2020quantum}. To this aim, there are already emerging several full ecosystems to exploit such methodologies and tools (e.g., \tild\cite{qpath,beisel2022quokka}), \new{and the compelling need for a structured discipline of Quantum Software Engineering is discussed by numerous authors (e.g., \tild\cite{barzen20222nd,piattini2021toward,scheerer2023experiences})}

\new{Jianjun Zhao defines in\tild\cite{zhao2020quantum} the term "Quantum Software Engineering" as}

\begin{quotation}
\noindent
\new{\textit{"The use of sound engineering principles for the development, operation, and maintenance of quantum software and the associated document to obtain economically quantum software that is reliable and works efficiently on quantum computers"}}
\end{quotation}
\medskip

\new{Highlighting the importance of applying "sound engineering principles" to the quantum software lifecycle, that its management must be  "economically" affordable, and that the resulting software must be "reliable" and must work "efficiently" on quantum computers.}

% The compelling need for a structured discipline of Quantum Software Engineering is discussed by numerous authors (e.g., \tild\cite{barzen20222nd,piattini2021toward,scheerer2023experiences}), as the way \textit{"to obtain economically quantum software that is reliable and works efficiently on quantum computers"}\tild\cite{zhao2020quantum}.

Some authors claim that Quantum Computing will lead to a new "Golden Age" of Software Engineering. They believe that \textit{"Software Engineering has built up a broad knowledge base, and has learnt many lessons that should be applied to the production of quantum software. The new quantum software engineering field needs to be considered as the application or adaptation of the well-known methods, techniques, and practices of software engineering. At the same time, however, new methods and techniques will be defined specifically for quantum software production"}\tild\cite{piattini2021quantum}.

These strong beliefs gather a large support all around the globe. As an example, "The Talavera Manifesto for Quantum Software Engineering and Programming"\tild\cite{talaveramanifesto}, a document that summarises the principles and commitments for Quantum Software Engineering, and that is considered a milestone in the (even if brief) history of QSE\tild\cite{de2023quantum}, has already been signed by more than 200 researchers and practitioners from more than 20 countries\footurl{https://www.aquantum.es/endorsers/}.

\section{Quantum Software Engineering}
\label{sec:qse}

%The history of Quantum Software Engineering (QSE) is pretty recent. The term \textit{"Quantum Software Engineering"} first appeared in 2002 in the "Grand Challenge for Computing Research"\tild\cite{grandchallenge} by John Clark and Susan Stepney. The challenge concerned \textit{"the development of a full discipline of Quantum Software Engineering, ready to exploit the full potential of commercial quantum computer hardware, once it arrives"}. The challenge identified also four different research lines, viz., Foundations, Languages and Compilers, Methods and Tools and, Novel Quantum Possibilities. 

The history of Quantum Software Engineering (QSE) is pretty recent. The term \textit{"Quantum Software Engineering"} first appeared in 2002 in the "Grand Challenge for Computing Research"\tild\cite{grandchallenge} by John Clark and Susan Stepney, in which the authors identify four different potential research lines, viz., Foundations, Languages and Compilers, Methods and Tools and, Novel Quantum Possibilities. 

As for the foundational aspects, the authors highlighted the need to further develop and investigate the concepts of \textit{Universal Turing Machine} and \textit{Quantum Algorithmic Complexity} as well as new models of quantum computations above the level of unitary matrices and gates.
Strictly linked with the new quantum computational models the authors discussed the need to determine new fundamental building blocks of quantum programming for assembly, high-level and specification languages, and the corresponding quantum compilers.
Such progress should be accompanied by new architectures and debugging and testing techniques, as well as more powerful simulators and visualisation techniques.
Finally, the authors recommended investigating the effects of the pure random generation and entanglement capabilities offered by quantum computers and how they can help to produce new algorithms and protocols.

%Moving forward, in 2020, "The Talavera Manifesto for Quantum Software Engineering and Programming"\tild\cite{talaveramanifesto} was presented as the resulting effort of academia and industry practitioners who joined at the \textit{first International Workshop on QuANtum SoftWare Engineering \& pRogramming}, collecting \textit{"some principles and commitments about the quantum software engineering and programming field, as well as some calls for action"}.

Moving forward, in 2020, "The Talavera Manifesto for Quantum Software Engineering and Programming"\tild\cite{talaveramanifesto} was presented as the resulting effort of academia and industry practitioners who joined at the \textit{first International Workshop on QuANtum SoftWare Engineering \& pRogramming}. 
The Manifesto discusses how QSE should:
\begin{enumerate}
    \item be agnostic regarding quantum programming languages and technologies,
    \item embrace the coexistence of classical and quantum computing,
    \item support the management of quantum software development projects,
    \item consider the evolution of quantum software,
    \item aim at delivering quantum programs with desirable zero defects,
    \item promote quantum software reuse,
    \item address security and privacy by design, and
    \item cover the governance and management of software.
\end{enumerate}

The Manifesto also links Quantum Software Engineering with classical Software Engineering suggesting how a quantum approach to Software Engineering should \textit{"take care of producing quantum software by applying knowledge and lessons learned from the software engineering field. This implies applying or adapting the existing software engineering processes, methods, techniques, practices and principles for the development of quantum software (or it may imply creating new ones)"}.

The same approach is also suggested in\tild\cite{serrano2022quantum}. The authors discuss that \textit{"the new quantum software engineering field needs to be considered as the application or adaptation of the well-known methods, techniques, and practices of software engineering. Some techniques can be used just as they are in classical computing. At the same time, however, new methods and techniques will be defined specifically for quantum software production"}.

Luis Barbosa, in his position paper\tild\cite{barbosa2020software}, identifies four main issues to a scientific rigorously Quantum Software Engineering discipline, viz., investigate appropriate \textit{semantic structures} capable of managing classical controls and quantum data, develop an \textit{algorithmic calculus} for the systematic derivation of quantum programs in a compositional way, seek a new family of \textit{dynamic logics} to support the formulation of contract for quantum programs and their compositional verification and, finally, design a \textit{framework for coordination} of orchestrated quantum computation systems. In the same work, Barbosa discusses also three research directions from a formal methods point of view, namely, the study of quantum \textit{models}, \textit{architectures} and, \textit{properties} with the ultimate goal of developing a \textit{"a mathematically based approach, able to conceptualise, and predict behaviour, and to provide a rich, formal framework for specifying, developing and verifying quantum algorithms"}.

In\tild\cite{piattini2021toward}, the authors identify software design, software construction, software testing, software maintenance, and software
quality as the main SWEBOK (Software Engineering Body of Knowledge)\tild\cite{SWEBOK2014} areas that will be heavily influenced by quantum computing, followed by software requirements, software engineering process, software engineering models and methods, and computing foundations. Furthermore, the authors discuss some promising research lines, viz., design of quantum hybrid systems, quantum program testing, quality assurance, and re-engineering and modernisation.

In\tild\cite{moguel2020roadmap}, the authors link the current state of Quantum Computing with that of classical computing in the late fifties and early sixties\tild\cite{10.1145/355604.361591,10.1145/800228.806933}, by considering different challenges and problems that researchers had to face in those periods and that today they reappeared in their quantum version, e.g., the hardware cost, their limited availability and limited power, the difficulty of operating quantum computers, their sensitivity and small reliability, the limited portability of the programs. On the basis of such considerations and the computer science history\tild\cite{mahoney1988history}, the authors warn that \textit{"all of the above [considerations] can lead one to assume that we are on the verge of a potential Quantum Software Crisis. Software Engineering must pay attention to these signals in order to anticipate it"}. At the same time, they believe that starting from this analogy and, \textit{"analyzing the advances and the lessons learned in the field of Software Engineering in the last 60 years, raises the directions that could help to develop the future Quantum Software Engineering"}. The authors outline also three possible directions for QSE, viz., investigating new quantum software \textit{processes} and \textit{methodologies}, design of new \textit{abstractions} for quantum software, and development of quantum \textit{structured programming}.

\new{One of the main comprehensive surveys on the field of QSE\tild\cite{zhao2020quantum}, was published by Jianjun Zhao in 2020. The "quantum software lifecycle" is a pivotal point in Zhao's work and after a careful discussion and analysis a first systemic, sequential model is proposed. Zhao's model is a five-step lifecycle comprising, namely, requirements analysis, software design, implementation, testing, and maintenance.}

In the same year, Weder \textit{et al.} proposed a quantum software lifecycle model too\tild\cite{weder2020quantum}. Their proposal, differently from Zhao's, starts by considering how to separate a problem's classical and quantum parts and is a ten steps process designed for applications during the NISQ era and, thus, incorporates phases such as data preparation, oracle expansion, and mitigation of readout-errors. Other steps are related to hardware-independent and dependent optimisation and the selection of a suitable quantum computer.

In\tild\cite{scheerer2023experiences}, the authors present a five-step quantum development model, based on their experiences and findings in the development of three Variational Quantum Algorithms (VQA)\tild\cite{vqa} for two industrial use cases (i.e., route planning and optimisation). The proposed phases are the following, \textit{problem definition}, \textit{quantum algorithm selection}, \textit{implementation}, \textit{fine-tuning}, and \textit{postprocessing}. The authors also noticed a high quantity of (even difficult) design decisions to be taken during each phase, most of them related to the problems induced by NISQ devices. They also highlight the need for experimental approaches to support the design decision process and underline the importance of \textit{Model-driven Software Development}\tild\cite{brambilla2017model} in quantum computing. In\tild\cite{9474563} too, Gemeinhardt \textit{et al.} support the study of \textit{Model-Driven Engineering} for quantum technologies, arguing its utility in easing the development of software systems.

With the aim of understanding what are the challenges and opportunities of quantum computing facing the software engineering community, El Aoun \textit{et al.} carried out an empirical study on Stack Exchange Forums and GitHub Issues to investigate the QSE-related challenges perceived by developers\tild\cite{DBLP:conf/icsm/aounLKO21}. They discovered that some of the challenges faced by quantum developers are the same present in classical software developments (e.g., dependency management). Anyway, quantum development also presents quantum-specific challenges (e.g., the interpretation of quantum programs' output). The authors also identify different areas requiring attention (e.g., learning resources both on practical and theoretical aspects of quantum computing, error management, and production of tools to support the development). 

With a similar approach, De Stefano \textit{et al.} mined GitHub repositories employing quantum framework to understand the most commonly used technologies and interviewed the contributors of such repositories to survey their opinions on the current adoption and challenges of quantum programming\tild\cite{de2022software}. They found out that quantum programming tools are mostly used for personal study purposes and most GitHub contributors on quantum computing work in research and framework repositories. They also discovered that most challenges are related to the understanding of quantum programs, the complexities associated with establishing hardware and software infrastructures, issues pertaining to implementation and code quality, the difficult task of building a quantum developer community, and the existing shortcomings in the realism of current quantum applications. They conclude how at the date quantum programming primarily serves didactic purposes or satisfies researchers' curiosity in experimenting with quantum technologies and that challenges are not only related to quantum development but also socio-technical considerations.

Focusing on quantum tools, Serrano \textit{et al.} reviewed main quantum software components and platforms\tild\cite{serrano2022quantumsoftware}. They discovered that most of the available quantum tools lack standard classical features such as project support, software governance, and a focus on software quality and evolution.
Another systematic survey on quantum tools, frameworks, and platforms is presented in\tild\cite{upama2022evolution}.

With a systematic study of broader scope, De Stefano \textit{et al.} surveyed existing literature on QSE\tild\cite{de2023quantum}. \new{They discovered that current QSE research \textit{"has primarily focused on software testing with little attention given to other topics, such as software engineering management"}.} Moreover, most papers proposed solutions and techniques or reported empirical findings and positions. They also propose to set the "official" establishment of QSE in 2020 with the Talavera Manifesto, with first publications, however, dating back to 2018.

\new{In the realm of \textit{quantum software testing}, researchers have grappled with the unique challenges posed by the principles of quantum mechanics. As highlighted by Miranskyy et al. \cite{miranskyy2019testing, miranskyy2021testing}, the very act of observing a quantum computation unavoidably alters its state, rendering traditional interactive debugging impractical. Consequently, a shift towards methodologies such as black-box testing or the judicious use of quantum simulators becomes imperative. Quantum simulators offer the advantage of observing qubit states without perturbation, allowing for more effective testing strategies.}

\new{Addressing these challenges, in \cite{garcia2023quantum} the authors delve into potential approaches for quantum software testing. Their exploration encompasses statistical techniques tailored to the stochastic nature of quantum physics. Leveraging statistical proof rules \cite{feng2007proof} and assertions \cite{yang2017proceedings}, they propose strategies for both the verification and testing phases. 
In the pursuit of verification, discussions revolve around quantum adaptations of Hoare logic \cite{miranskyy2020your}, a formal system for reasoning about the correctness of computer programs. Additionally, the authors illustrate how the concept of quantum reversibility, as elucidated by Patel et al. \cite{patel2004fault} and Zamani et al. \cite{zamani2012ping}, can be harnessed for verification purposes, further enhancing the reliability of quantum programs.}

\section{The Future of Quantum Software Engineering}

The existing body of literature indicates a growing and pressing demand for the creation of a novel Software Engineering paradigm tailored for Quantum Computing. This need becomes even more pronounced as the quantity and quality of quantum computers continue to advance. This approach should be adept at addressing problems by either building upon and potentially enhancing classical techniques and tools or by pioneering entirely new solutions, facing challenges unique to the quantum domain. This new Quantum Software Engineering must possess the capability to address the challenges that the next generation of quantum developers will encounter. Notably, the upcoming cohort of quantum developers will comprise individuals who are not solely specialists in the field with extensive experience in quantum mechanics and computing. Thus, the new Quantum Software Engineering proposal must address also such a heterogeneous user base. 

\new{Numerous papers have explored intriguing and potentially transformative avenues for future research in Quantum Software Engineering (QSE) \cite{carleton2021architecting,moguel2020roadmap,zhao2020quantum,de2023quantum,DBLP:conf/icsm/aounLKO21,de2022software}. While these directions hold considerable merit and interest, we have chosen not to reiterate them in this manuscript. Instead, we decided to delve into uncharted territories, proposing research directions that we anticipate harbouring significant potential yet remain, to the best of our knowledge, barely addressed in the existing literature. Herein, we present promising research avenues and challenges within QSE, anticipating their escalating importance in the near future and envisaging their potential to yield groundbreaking discoveries:}

\begin{enumerate}
    \item \new{\textit{Language abstractions.} Developing Quantum algorithms remains a nuanced art rather than a streamlined engineering process. Presently, quantum abstractions mirror the early stages of classical computing, where each individual (qu)bit and gate requires meticulous management. Crafting intricate quantum programs thus poses formidable challenges. However, with the development of higher-level language abstractions and foundational quantum primitives, developers can potentially transcend the burdensome intricacies of low-level matrix operations. This shift promises to bring quantum development closer to the intuitive and efficient coding practices prevalent in modern classical programming paradigms\tild\cite{chong2017programming,furntratt2023towards,bichsel2020silq}.} 
    \item \new{\textit{Quantum software debugging and visualisation.} Debugging and visualising quantum software pose unique challenges due to the inherent nature of quantum computation. Unlike classical computing, observing the state of a quantum computation inevitably disrupts its execution, complicating the debugging process on real quantum hardware. Fortunately, quantum simulators offer a workaround, enabling observation of qubit states without perturbation. By refining debugging techniques tailored to quantum environments, we can enhance the quality and ease of developing quantum programs. Moreover, improving visualisation techniques for quantum computation holds promise in enhancing both comprehension and debugging capabilities. Clear visual representations of quantum processes can provide invaluable insights into program behaviour and facilitate the identification of errors. By investing in the advancement of quantum software debugging and visualisation tools, we can significantly accelerate progress in quantum computing\tild\cite{miranskyy2021testing,di2024need,sasakura2023potential}.}
    \item \new{\textit{Distributed quantum computations.} The current quantum computing landscape is characterised by a significant diversity in qubit implementations and quantum computer architectures, resulting in a broad spectrum of performance and qualitative attributes. Presently, quantum computers are predominantly perceived and utilised as individual monolithic entities. However, an alternative approach could involve distributing quantum computations across multiple quantum computers, capitalising on and leveraging the existing heterogeneity rather than perceiving it as a limitation.  This paradigm shift opens doors to novel strategies in quantum computation. By harnessing the diverse capabilities of various quantum computing platforms, we can envision a distributed computing framework where tasks are intelligently allocated across a network (either classical or quantum) of interconnected quantum devices. This distributed approach not only mitigates the limitations imposed by individual quantum computers but also unlocks synergistic potentials arising from their collective strengths. Embracing this heterogeneity fosters a more robust and scalable quantum computing ecosystem, paving the way for collaborative problem-solving on a scale previously unattainable\tild\cite{cuomo2020towards,bisicchia2023dispatching,bisicchia2023distributing}.}
\end{enumerate}

\ifCLASSOPTIONcaptionsoff
  \newpage
\fi



\bibliographystyle{IEEEtran}
\bibliography{biblio}
\end{document}